\documentstyle[12pt,epsfig,amsfonts,here]{article}

\def\Journal#1#2#3#4{{#1} {\bf #2}, #3 (#4)}

\def\NPB{{\em Nucl. Phys.} B}
\def\PLB{{\em Phys. Lett.}  B}
\def\PRL{\em Phys. Rev. Lett.}
\def\PRD{{\em Phys. Rev.} D}
\def\ZPC{{\em Z. Phys.} C}
\def\EPG{{\em Eur.Phys.J.} C}
\def\IJMP{\em Int. Jour. of Mod. Phys.}

\def\be{\begin{equation}}
\def\ee{\end{equation}}
\def\bea{\begin{eqnarray}}
\def\eea{\end{eqnarray}}

\begin{document}
\begin{center}
{\large\bf THE FIRST THREE POMERONS...}
\vskip 1.cm
{V. A. PETROV$^{a,}$\footnote{\it E-mail: petrov@mx.ihep.su}
 and A. V. PROKUDIN$^{a,b,}$\footnote{\it E-mail: prokudin@to.infn.it}}
\vskip 0.5cm
{\small\it
(a) Institute For High Energy Physics,\\ 
142281 Protvino,  RUSSIA}
\vskip 0.2cm 

{\small\it
\vskip 0.2cm 
(b)  Dipartimento di Fisica Teorica,\\ 
Universit\`a Degli Studi Di Torino, \\
Via Pietro Giuria 1,
10125 Torino, \\ 
ITALY\\
and\\
Sezione INFN di Torino,\\
 ITALY\\}
\vskip 0.5cm


\parbox[t]{12.cm}{\footnotesize A model of a three Pomeron contribution to high energy 
elastic $pp$ and $\bar p p$ scattering is proposed. The data are well
described for all momenta ($0.01\le |t|\le 14.\; GeV^2$) and energies 
($8.\le\sqrt{s}\le 1800.\; GeV$) ($\chi^2/{\rm d.o.f.}=2.74$). The model 
predicts the appearance of two dips in the differential cross-section 
which will be measured at LHC. The parameters of the Pomeron trajectories are:
 \\
$\alpha(0)_{{\Bbb P}_1}=1.058,\;\;\alpha'(0)_{{\Bbb P}_1}=0.560\;(GeV^{-2});$ \\
$\alpha(0)_{{\Bbb P}_2}=1.167,\;\;\alpha'(0)_{{\Bbb P}_2}=0.273\;(GeV^{-2});$ \\
$\alpha(0)_{{\Bbb P}_3}=1.203,\;\;\alpha'(0)_{{\Bbb P}_3}=0.094\;(GeV^{-2}).$ \\
}
\end{center}

\section{INTRODUCTION}

Fervently awaited high-energy collisions at LHC will give an access 
not only to yet unexplored
small distances but also simultaneously to neither explored 
large distances~\cite{Petrov}.
Future measurements of total and elastic cross-sections 
at LHC~\cite{multipomeronFaus-Golfe}
tightly related to the latter domain 
naturally stimulate further searches
for new approaches to diffractive scattering at high energies.
 
Recently some models with multi-Pomeron structures were proposed
~\cite{{multipomeronNicolescu},{multipomeronKontros},
{multipomeronLandshoff}}. Some of them~\cite{multipomeronNicolescu},~\cite{multipomeronKontros} use Born amplitudes with two Pomerons as single 
~\cite{multipomeronNicolescu} or double poles~\cite{multipomeronKontros}. 
Formal violation of the Froissart-Martin bound in some of these models 
is considered as ``practically negligible"  though 
in terms of parial-wave amlitudes
unitarity violation is flagrant at present-day energies. Nonetheless a model
of such kind ~\cite{multipomeronLandshoff} based on the two-Pomeron 
approach shows quite a  good agreement with DIS data.

The eikonal models that are capable of describing the data for nonzero	transferred momenta are developed in Refs~\cite{multipomeronPredazzi},
~\cite{multipomeronPancheri}. In some cases a ``generalized eikonal 
representation" is used~\cite{multipomeronPredazzi} together with a 
dipole (monopole) Pomeron contibution, 
in the others the conventional eikonal is supplemented 
with a ``QCD motivated" part consisting of three terms~\cite{multipomeronPancheri}.
It is worth noticing that the two-Pomeron eikonal has been applied 
to the description of the data more than ten years ago 
(see, e.g., Ref.~\cite{multipomeronLikhoded}).

The very multiformity of the models hints that maybe the most 
general way to describe high-energy diffraction is just to admit 
an arbitrary number of Pomerons (i.e. all vacuum Regge-poles contributing 
non-negligibly at reasonably high energies. Roughly, they should have
intercepts not lower than $1$). On the one hand this seems not very economical.
But on the other hand we could argue that no basic 
principle forbids more than one single Pomeron. We could also add that in the 
perturbative framework the account of the renormalization group
leads presumably to converting of the fixed branch point (in the $J$-plane)
 into an infinite series of simple poles accumulating down to 1 
from some maximal value~\cite{Lipatov}. Unfortunately perturbative searches in this field
are far from being satisfactory from many viewpoints.

In this paper we would like to make a first step in realization 
of the above formulated hypothesis about many-Pomeron structure of the eikonal.
As it seems not possible to describe the data in the framework of
the eikonal approach with presence of one single pole Pomeron 
contribution~\cite{mypaper}, and the two-Pomeron option does not improve
quality of description drastically (more details are given in the text) 
it is fairly natural
to try the next, three-Pomeron, option for the eikonal.
We will see below that this choice appears rather lucky.

\section{THE MODEL}
Let us brifely outline the basic properties of our model. Unitarity condition:
$$
\Im{\rm m}\; T(s,\vec b) = \vert T(s,\vec b)\vert^2 + \eta (s,\vec b)\; ,
$$
where $T(s,\vec b)$ is the scattering amplitude in the impact
representation,
$\vec b$ is the impact parameter, $\eta (s,\vec b)$ is the
contribution of 
inelastic channels, implies the following eikonal form for the scattering amplitude $T(s,\vec b)$
\begin{equation}
T(s,\vec b)=\frac{e^{2i\delta (s,\vec b)}-1}{2i}\; ,
\label{eq:ampl}
\end{equation}
where $\delta (s,\vec b)$ is the
eikonal function. The unitarity condition in terms of the eikonal looks as follows
\begin{equation}
{\rm \Im m}\; \delta (s,\vec b) \ge 0, \; s>s_{\rm inel}\; .
\label{eq:euc}
\end{equation}

The eikonal function is assumed to have simple poles in the complex $J$-plane 
and the
corresponding Regge trajectories are normally being used in the linear 
approximation
\be
\alpha(t) = \alpha(0) + \alpha '(0)t\; .
\label{eq:rt}
\ee

Accordingly we get the following (modulo the signature factor) 
contribution to the
eikonal function in $t$-space (here $t$ is the momentum transfer)
\be
\displaystyle \hat \delta (s,t) = \frac{c}{s_0} \Big( \frac{s}{s_0} \Big)
^{\alpha(0)}e^{t\frac{\rho^2}{4}}\; ,
\label{eq:eikonalt}
\ee
where  
\be
\rho^2 = 4\alpha'(0) \ln\frac{s}{s_0}+r^2
\ee
\noindent
is referred to as the ``Reggeon radius''.

In order to relate $t$- and $b$-spaces one proceeds via Fourier-Bessel
transforms
\be
\begin{array}{r}
\displaystyle \hat f(t)= 4 \pi s\int_{0}^{\infty} db^2 J_0(b\sqrt{-t}) f(b)\; , \\
\\
\displaystyle f(b)= \frac{1}{16 \pi s}\int_{-\infty}^{0} dt J_0(b\sqrt{-t}) \hat f(t) \; .
\end{array}
\label{eq:fb}
\ee

Making use of Eq.~(\ref{eq:fb}) we obtain the following $b$-representation
of the eikonal function~(\ref{eq:eikonalt})
\be
\displaystyle \delta (s,b) = \frac{c}{s_0}
\Big(\frac{s}{s_0}\Big)^{\alpha(0)-1}\frac{e^{-\frac{b^2}{\rho^2}}}{4\pi
\rho^2}\; .
\label{eq:eikonalb}
\ee

For the cross-sections we use the following normalizations:
\be
\begin{array}{l}
\displaystyle \sigma_{tot} = \frac{1}{s} \Im {\rm m} T(s,t=0), \\
\\
\displaystyle \sigma_{elastic} = 4 \pi \int_{0}^{\infty}db^2 \vert T(s,b) \vert^2, \\
\\
\displaystyle \frac{d\sigma}{dt} = \frac{\vert T(s,t) \vert ^2}{16\pi s^2}, \\
\\
\displaystyle \rho =\frac{\Re e T(s,t=0)}{\Im m T(s,t=0)}.
\end{array}
\label{eq:norm}
\ee

In the present model we assume the following representation for the eikonal function:
\be
\delta_{pp}^{\bar p p}(s,b) = \delta^+_{{\Bbb P}_1}(s,b)+
\delta^+_{{\Bbb P}_2}(s,b)+
\delta^+_{{\Bbb P}_3}(s,b)
\mp \delta^-_{\Bbb
O}(s,b)+\delta^+_{
f}(s,b)\mp \delta^-_{\omega}(s,b),
\label{eq:modeleik}
\ee

here $\delta^+_{{\Bbb P}_{1,2,3}}(s,b)$ are Pomeron 
contributions. `$+$' denotes C even trajectories (the Pomeron trajectories have the following quantum numbers $0^+J^{++}$),
`$-$' denotes  C odd trajectories, $\delta^-_{\Bbb O}(s,b)$ is the 
Odderon contribution 
(the Odderon is the C odd partner of 
the Pomeron with quantum numbers $0^-J^{--}$);
$\delta^+_{ f}$, 
$\delta^-_{\omega}(s,b)$ are the contributions of secondary Reggeons, $f$
($C=+1$) and
$\omega$ ($C=-1$).

The form (\ref{eq:eikonalt}) is not compatible with analyticity
and crossing symmetry, which are easily restored by substitution 
$s \rightarrow s e^{-i\pi/2}$. We introduce a new dimensionless variable 
\be
\tilde s = \frac{s}{s_0}e^{-i\frac{\pi}{2}}\; ,
\ee
and obtain each $C+$ and $C-$ contribution
with its appropriate 
signature factor and the form:

\bea
\displaystyle \delta^+ (s,b)=i\frac{c}{s_0}
\tilde s^{\alpha(0)-1}\frac{e^{-\frac{b^2}{\rho^2}}}{4\pi \rho^2}\; , 
\label{eq:eikonalform} \\
\nonumber
\rho^2 = 4\alpha'(0) \ln\tilde s+r^2\; , \\
\nonumber
(C = +1)\; ;
\\ 
\displaystyle \delta^- (s,b)=\frac{c}{s_0}
\tilde s^{\alpha(0)-1}\frac{e^{-\frac{b^2}{\rho^2}}}{4\pi \rho^2}\; ,
\label{eq:eikonalform1}
\\
\nonumber
\rho^2 = 4\alpha'(0) \ln\tilde s+r^2\; , \\
\nonumber
(C = -1)\; .
\eea

The parameters of secondary Reggeon trajectories are fixed according to
the parameters obtained from a fit of the meson spectrum
~\cite{gd93}:
\be
\begin{array}{l}
\alpha_f(t) = 0.69+0.84 t\; , \\
\\
\alpha_\omega (t) = 0.47+0.93 t\; .
\end{array}
\ee

All the trajectories are taken in linear aproximation:
\be
\alpha_i(t) = \alpha_i(0)+\alpha_i'(0)t,\;(i={\Bbb P}_1,{\Bbb P}_2,
{\Bbb P}_3,{\Bbb O}).
\ee
Let us remark that all good fits require $\alpha_{\Bbb P} (0)-1\equiv 
\Delta_{\Bbb P}>0$ which means that Born amplitude will eventually exceed
the Froissart-Martin~\cite{fm} unitarity bound. This violation of 
unitarity is removed by all kinds of ``eikonalization''. Nevertheless, one
must take into account the following unitarity 
constrains~\cite{pomeronodderon}
\be
\alpha_{\Bbb P}(0) \ge \alpha_{\Bbb O}(0)\; {\rm and}\; 
\alpha'_{\Bbb P}(0) \ge \alpha'_{\Bbb O}(0)\; ,
\label{eq:uconstraints}
\ee
there $\Bbb P$ in this case is the leading Pomeron trajectory 
(the one with the highest intercept $\Delta_{\Bbb P}$).

\section{RESULTS}

We fitted the adjustable parameters over a set of 961 $pp$ and $\bar p p$ 
data of both forward observables (total cross-sections $\sigma_{tot}$,
and $\rho$ -- ratios of real to imaginary part of the amplitude) in the
range $8.\le\sqrt{s}\le 1800.\; GeV$ and angular distributions 
($\frac{d\sigma}{dt}$) in the ranges $23.\le\sqrt{s}\le 1800.\; GeV$,
$0.01\le |t|\le 14.\; GeV^2$. 

Having used 20 adjustable parameters we achieved $\chi^2/{\rm d.o.f.}=2.74$. 
The parameters are presented in Table~\ref{tab:1} (all the errors are obtained 
according to MINUIT output).
\begin{table}[H]
\begin{center}
\begin{tabular}{|l|l|l|l|}
\hline
& {\bf Pomeron$_{\bf 1}$} & & {\bf $\bf f$-Reggeon}  \\
\hline
$\Delta_{{\Bbb P}_1}$ & $0.0578\pm0.0020$ & $\Delta_{f}$ & $-0.31$ (FIXED) \\
$c_{{\Bbb P}_1}$ & $53.007\pm0.795$  & $c_{f}$ & $191.69\pm2.12$            \\
$\alpha'_{{\Bbb P}_1}$& $0.5596\pm0.0078\;(GeV^{-2})$  & $\alpha'_{f}$& $0.84\;(GeV^{-2})$ (FIXED)    \\
$r^2_{{\Bbb P}_1}$& $6.3096\pm0.2522\;(GeV^{-2})$&$r^2_{f}$ & $31.593\pm1.099\;(GeV^{-2})$  \\
\hline
& {\bf Pomeron$_{\bf 2}$}& & {\bf $\bf \omega$-Reggeon}  \\
\hline
$\Delta_{{\Bbb P}_2}$ & $0.1669\pm0.0012$  & $\Delta_{\omega}$ & $-0.53$ (FIXED)   \\
$c_{{\Bbb P}_2}$ & $  9.6762\pm0.1600$    & $c_{\omega}$ & $-174.18\pm2.72$           \\
$\alpha'_{{\Bbb P}_2}$& $0.2733\pm0.0056\;(GeV^{-2})$    & $\alpha'_{\omega}$& $0.93\;(GeV^{-2})$ (FIXED)     \\
$r^2_{{\Bbb P}_2}$& $3.1097\pm0.1817\;(GeV^{-2})$ &$r^2_{\omega}$ & $7.467\pm1.083\;(GeV^{-2})$  \\
\hline
& {\bf Pomeron$_{\bf 3}$}& & \\
\hline
$\Delta_{{\Bbb P}_3}$ & $0.2032\pm0.0041$  &$s_0$& $1.0\;(GeV^2)$ (FIXED)   \\
$c_{{\Bbb P}_3}$ & $1.6654\pm0.0669$       & &       \\
$\alpha'_{{\Bbb P}_3}$& $0.0937\pm0.0029\;(GeV^{-2})$   & &     \\
$r^2_{{\Bbb P}_3}$& $2.4771\pm0.0964\;(GeV^{-2})$ & & \\
\hline
& {\bf Odderon}& & \\
\hline
$\Delta_{{\Bbb O}}$ & $0.19200\pm0.0025$  & &  \\
$c_{{\Bbb O}}$ & $0.0166\pm0.0022$      & &        \\
$\alpha'_{{\Bbb O}}$& $0.048\pm0.0027\;(GeV^{-2})$   & &     \\
$r^2_{{\Bbb O}}$& $0.1398\pm0.0570\;(GeV^{-2})$ & & \\
\hline
\end{tabular}
\end{center}
\caption{Parameters obtained by fitting to the data. \label{tab:1}}
\end{table}

In order to estimate the quality of the description,
we have calculated partial $\chi^2$ over all sets of data used in the fit.
This $\chi^2$ is calculated using the following formula:
\be
\chi^2 = \sum_{n=1}^{ntot}\frac{(\sigma_{theory}(n)-\sigma_{exp}(n))^2}
{(\Delta(\sigma_{exp}(n)))^2}\; ,
\ee
where $ntot$ is the number of data in the set, $\sigma_{exp}$ is the 
experimental value of the quantity that is
described,  $\sigma_{theory}$ is our prediction for this quantity, and $\Delta(\sigma_{exp}(n))$ is the experimental uncertainty.

 The partial $\chi^2$ may be found in Table~\ref{tab:2}. 
\begin{table}[H]
\begin{center}
\begin{tabular}{|l|l|l|l|}
\hline
&{\bf Set of data} & {\bf Number of points, ntot} & 
{\bf $\chi^2/{\rm ntot}$} \\
\hline
& & & \\
1& $\sigma_{total}^{\bar p p}$ & 42  &  2.3524 \\
2& $\sigma_{total}^{p p}$ & 50  &  0.6309  \\
3& $\rho^{\bar p p}$ & 11  & 0.6942  \\
4& $\rho^{p p}$ & 36  & 1.9075   \\ 
5& $\frac{d\sigma}{dt}^{\bar p p},\;\sqrt{s}=31.\:(GeV)$ & 22  &  3.3688 \\
6& $\frac{d\sigma}{dt}^{\bar p p},\;\sqrt{s}=53.\:(GeV)$ & 52  &  8.5457  \\
7& $\frac{d\sigma}{dt}^{\bar p p},\;\sqrt{s}=62.\:(GeV)$ & 23  &  1.8524 \\
8& $\frac{d\sigma}{dt}^{\bar p p},\;\sqrt{s}=546.\:(GeV)$ & 78  &  3.8425  \\
9& $\frac{d\sigma}{dt}^{\bar p p},\;\sqrt{s}=630.\:(GeV)$ & 19  & 9.9273  \\
10& $\frac{d\sigma}{dt}^{\bar p p},\;\sqrt{s}=1800.\:(GeV)$ & 51  & 1.3741  \\
11& $\frac{d\sigma}{dt}^{p p},\;\sqrt{s}=23.5\:(GeV)$ & 105  & 2.2491  \\
12& $\frac{d\sigma}{dt}^{p p},\;\sqrt{s}=27.43\:(GeV)$ & 39  & 1.8929  \\
13& $\frac{d\sigma}{dt}^{p p},\;\sqrt{s}=30.7\:(GeV)$ & 92  &  4.4559 \\
14& $\frac{d\sigma}{dt}^{p p},\;\sqrt{s}=44.64\:(GeV)$ & 97  & 1.5748  \\
15& $\frac{d\sigma}{dt}^{p p},\;\sqrt{s}=52.8\:(GeV)$ & 93  &  2.0956 \\
16& $\frac{d\sigma}{dt}^{p p},\;\sqrt{s}=62.\:(GeV)$ & 151  &  2.4272 \\
\hline
 & {\bf Number of parameters}&{\bf Total number of points}& 
{$\chi^2/{\rm d.o.f.}$} \\
\hline
& 20 & 961 &  2.7441 \\ 
\hline 
\end{tabular}
\end{center}
\caption{Partial $\chi^2$. \label{tab:2}}
\end{table}

Some of these $\chi^2$s are high (for instance those for differential cross sections at $\sqrt{s}=53,\; 630\; GeV$). It reflects the fact that we did not make use of systematical errors for these sets of data which can be as high as $30\%$.  

\begin{figure}[H]
{\vspace*{ -2cm} \epsfxsize=140mm \epsffile{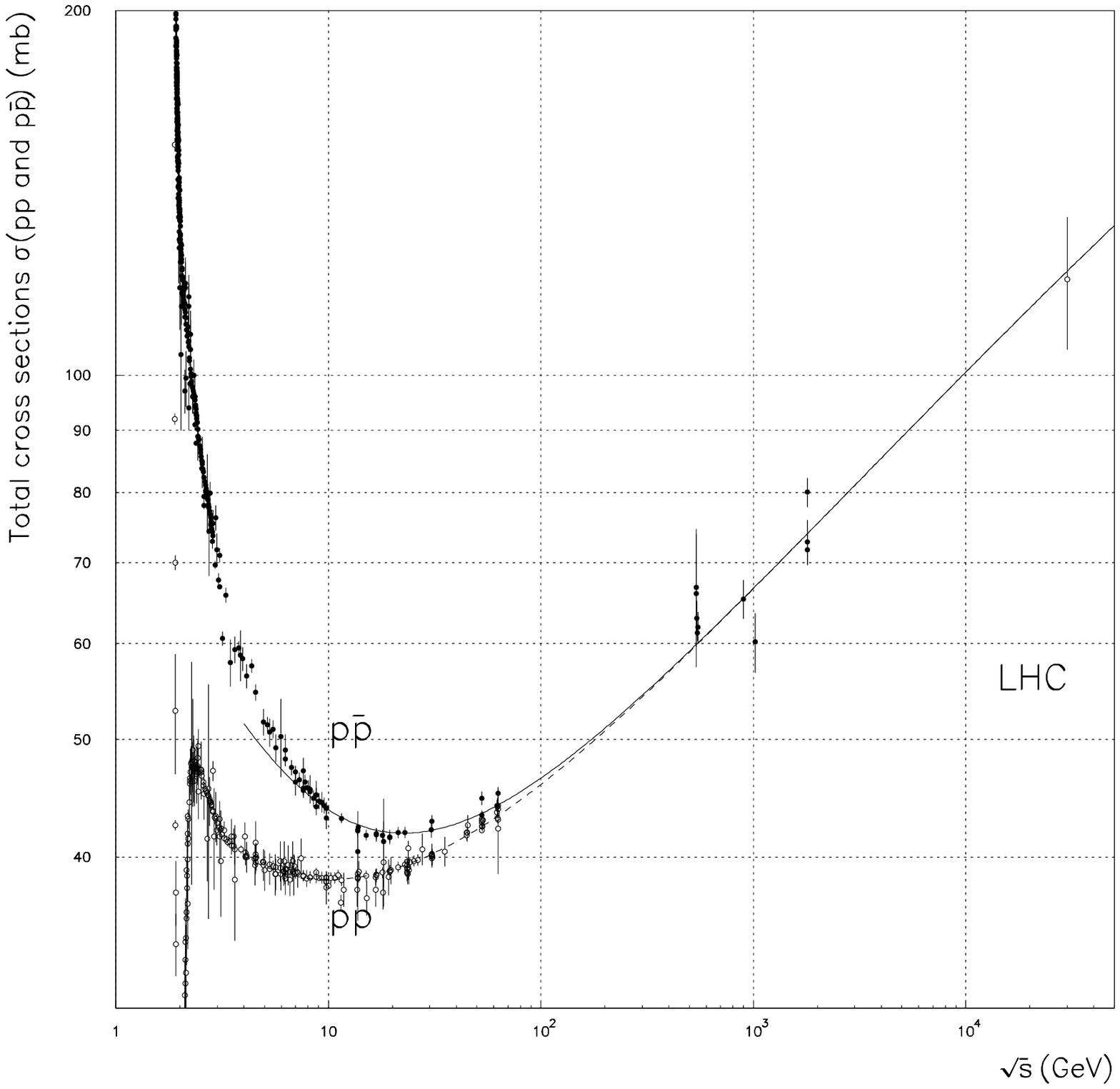}}
\vskip -3.cm
\caption{Total cross sections of $pp$ scattering 
(hollow circles)  and $\bar p p$ scattering (full circles)
and curves corresponding to their description in the present model.
\label{fig:tot}
}
\end{figure}

\begin{figure}[H]
{\vspace*{ -2cm} \epsfxsize=140mm \epsffile{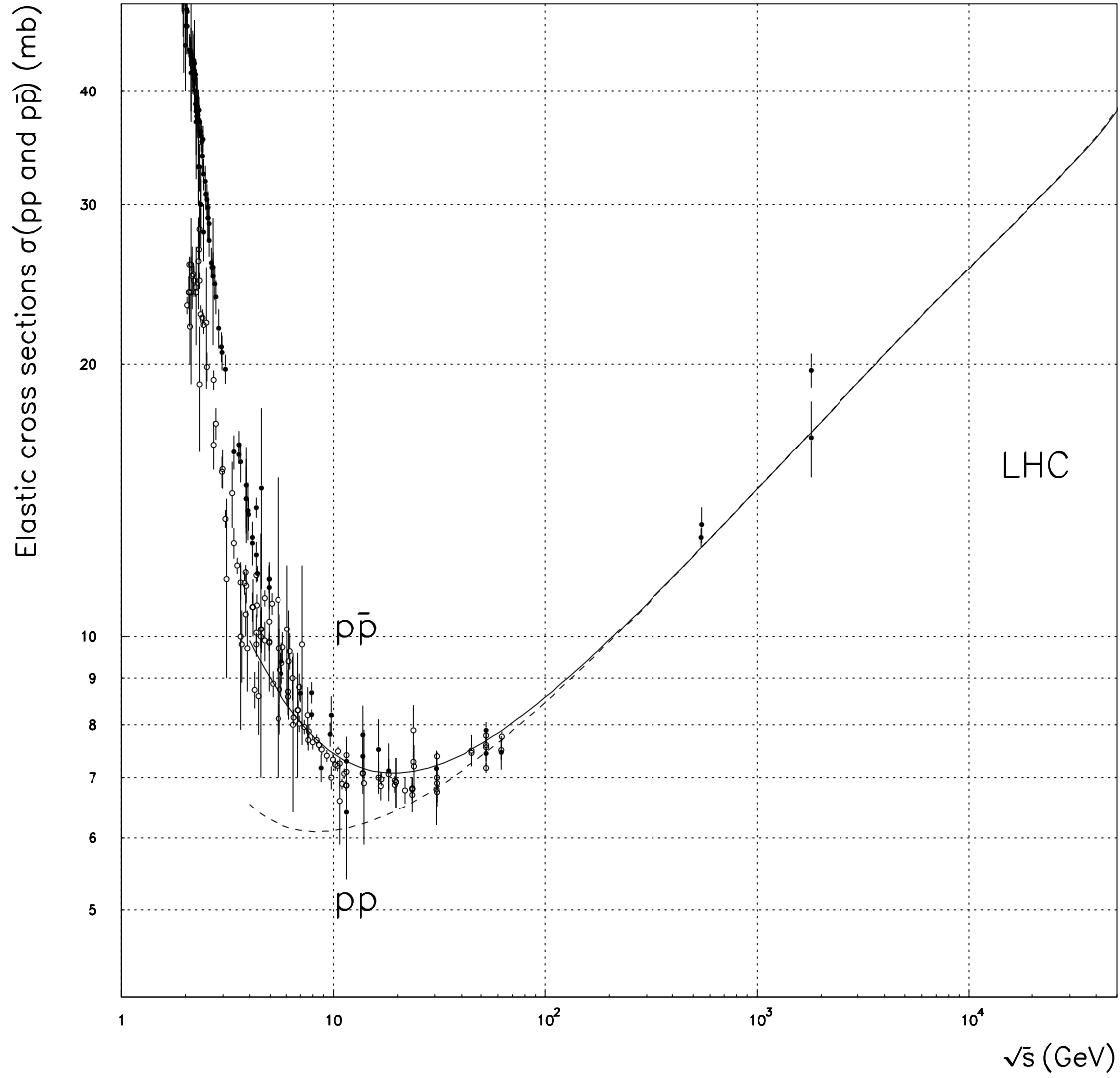}}
\vskip -3.cm
\caption{Elastic cross sections of $pp$ scattering 
(hollow circles)  and $\bar p p$ scattering (full circles)
and curves corresponding to their description in the present model.
These sets of data are not included in the fit.
\label{fig:elastic}
}
\end{figure}

\begin{figure}[H]
{\vspace*{ -2cm} \epsfxsize=140mm \epsffile{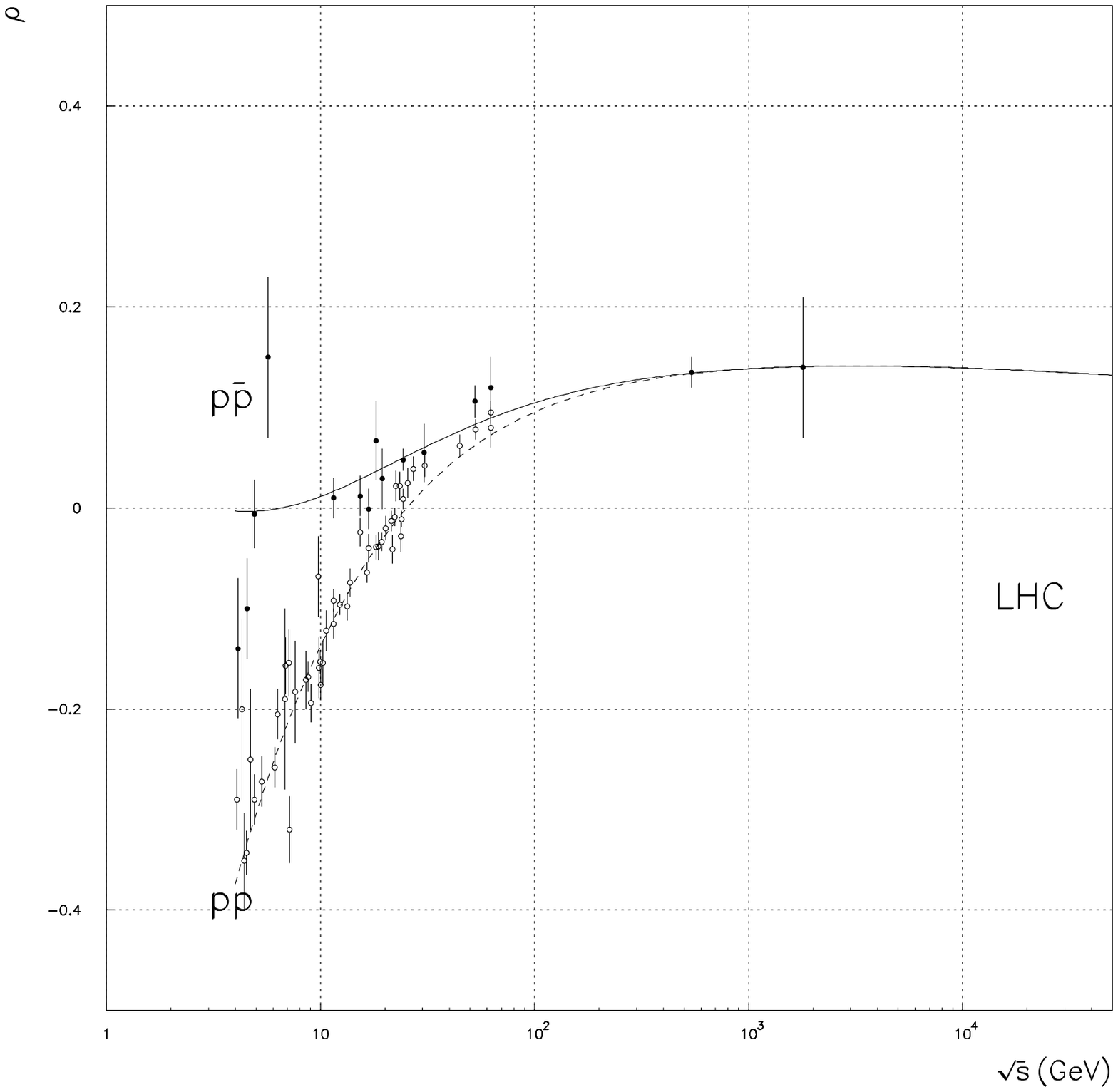}}
\vskip -3.cm
\caption{Ratios of the real to the imaginary part of the forward $pp$ scattering amplitude 
(hollow circles)  and $\bar p p$ scattering  amplitude (full circles)
and curves corresponding to their description in the present model.
\label{fig:reim}}
\end{figure}

\begin{figure}[H]
{\vspace*{ -2cm} \epsfxsize=140mm \epsffile{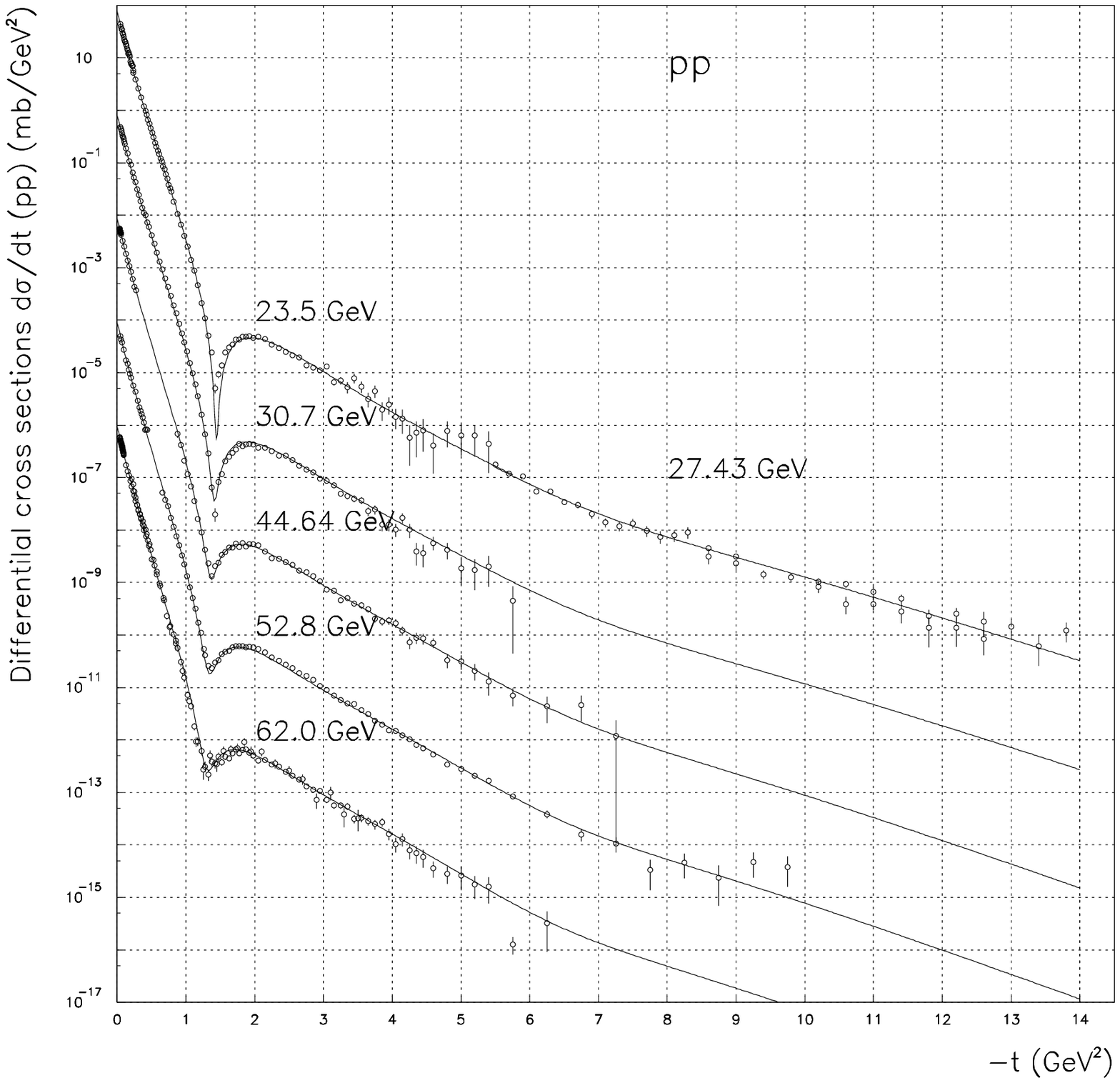}}
\vskip -3.cm
\caption{Differential cross-sections for $pp$ scattering
and curves corresponding to their description in the present model. 
A $10^{-2}$ factor between each successive set of data is omitted. 
\label{fig:ppdif}}
\end{figure}

\begin{figure}[H]
{\vspace*{ -2cm} \epsfxsize=140mm \epsffile{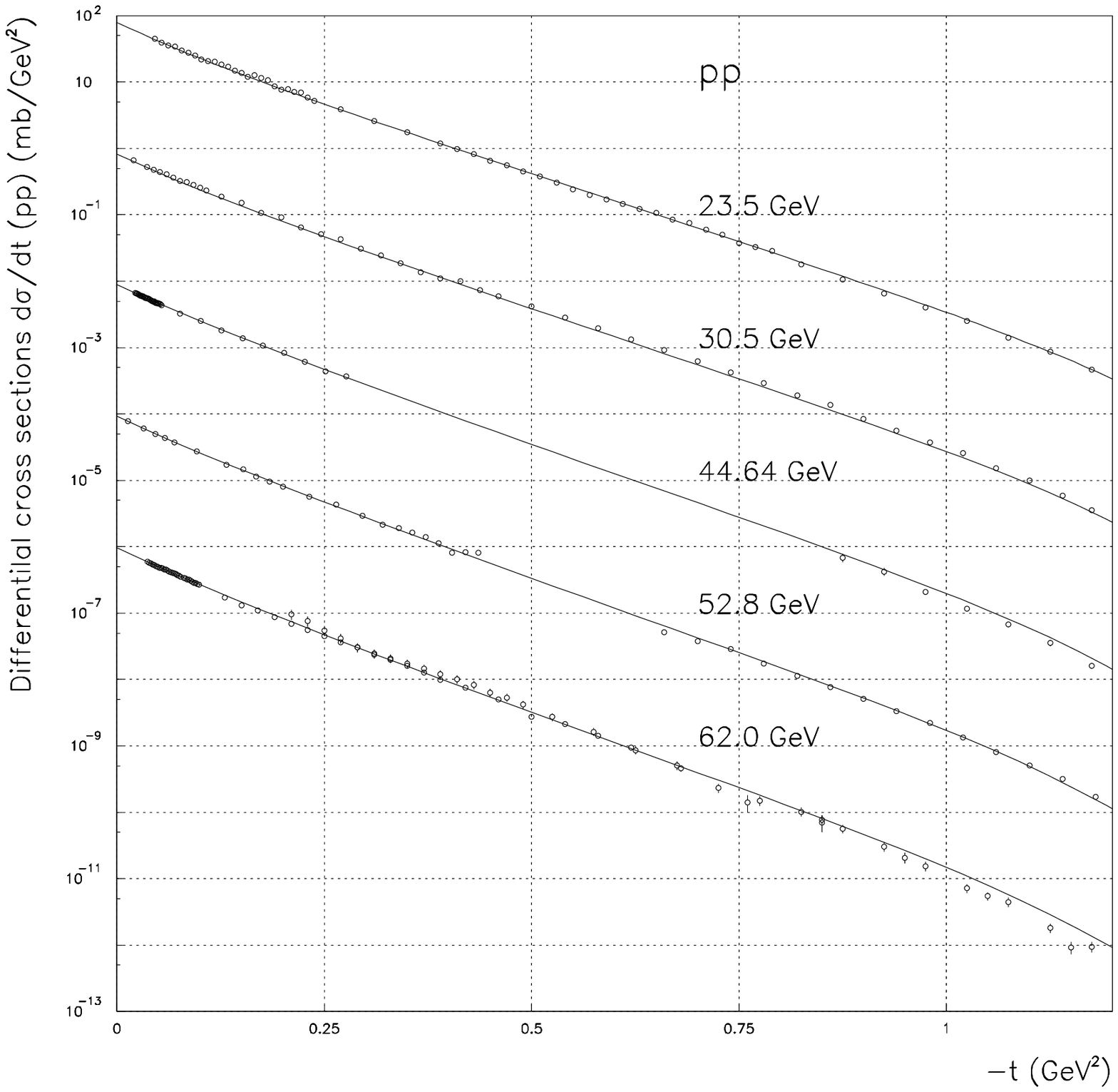}}
\vskip -3.cm
\caption{Differential cross-sections for $pp$ scattering in the region of 
small momenta $0.01\le|t|\le 1.2\; GeV^2$ and curves corresponding to their description in present the model.   
A $10^{-2}$ factor between each successive set of data is omitted. 
\label{fig:ppdif1}}
\end{figure}

\begin{figure}[H]
{\vspace*{ -2cm} \epsfxsize=140mm \epsffile{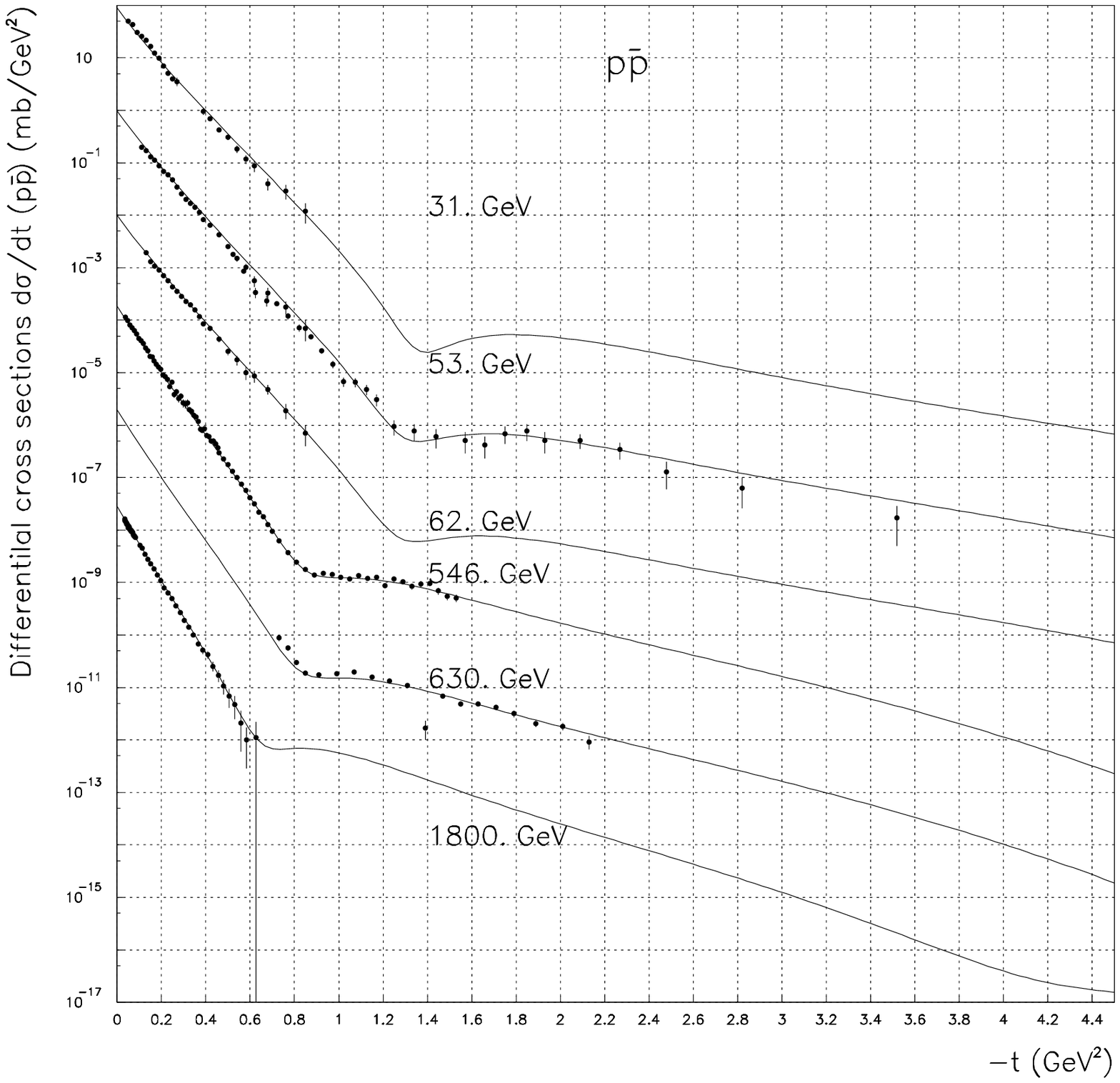}}
\vskip -3.cm
\caption{Differential cross-sections for $\bar p p$ scattering
and curves corresponding to their description in the present model. 
A $10^{-2}$ factor between each successive set of data is omitted. 
\label{fig:difpbarp}}
\end{figure}

\begin{figure}[H]
{\vspace*{ -2cm} \epsfxsize=140mm \epsffile{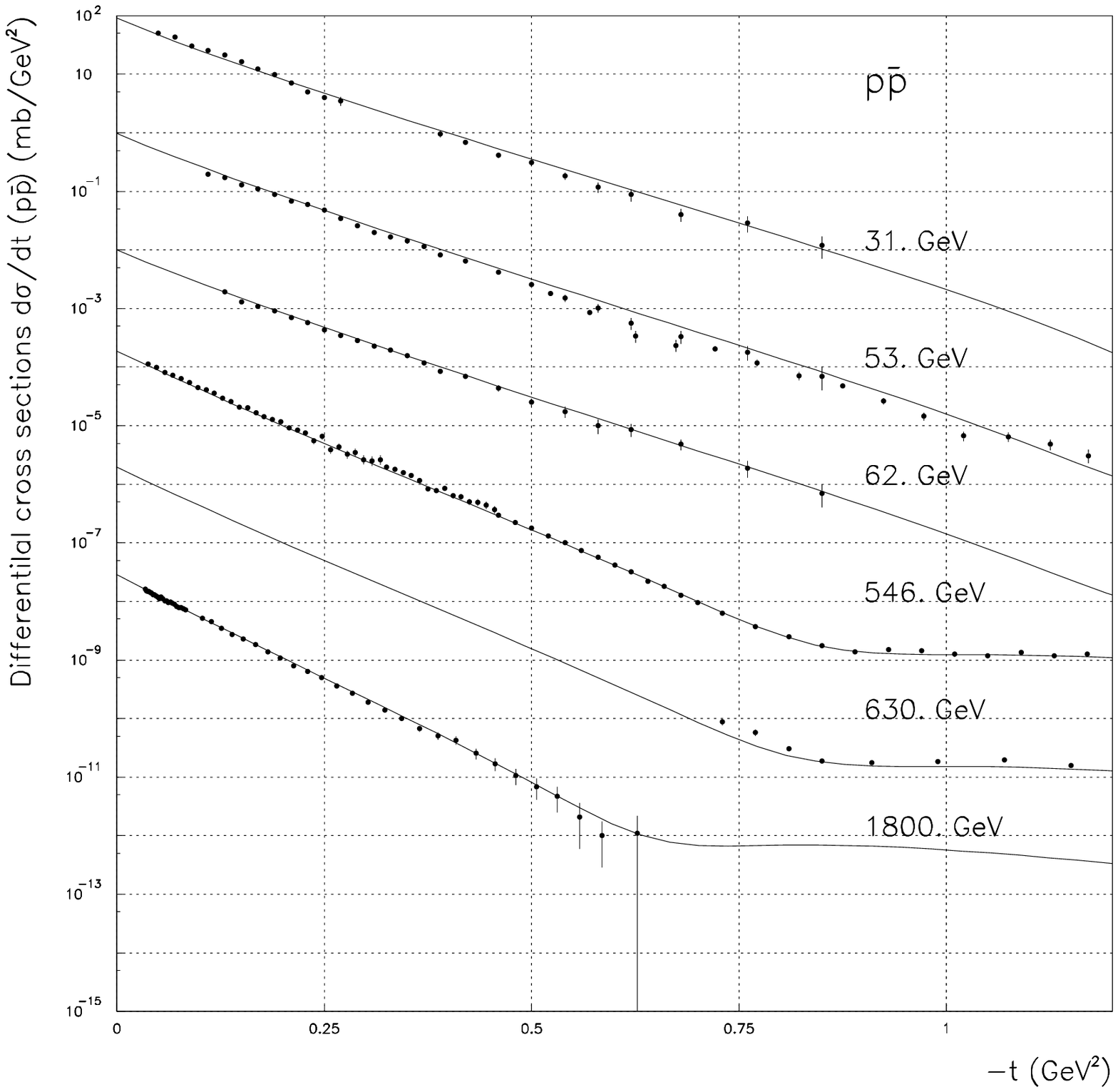}}
\vskip -3.cm
\caption{Differential cross-sections for $pp$ scattering in the region of 
small momenta $0.01\le|t|\le 1.2\; GeV^2$
and curves corresponding to their description in the present model. 
A $10^{-2}$ factor between each successive set of data is omitted. 
\label{fig:difpbarp1}}
\end{figure}

\begin{figure}[H]
{\vspace*{ -2cm} \epsfxsize=140mm \epsffile{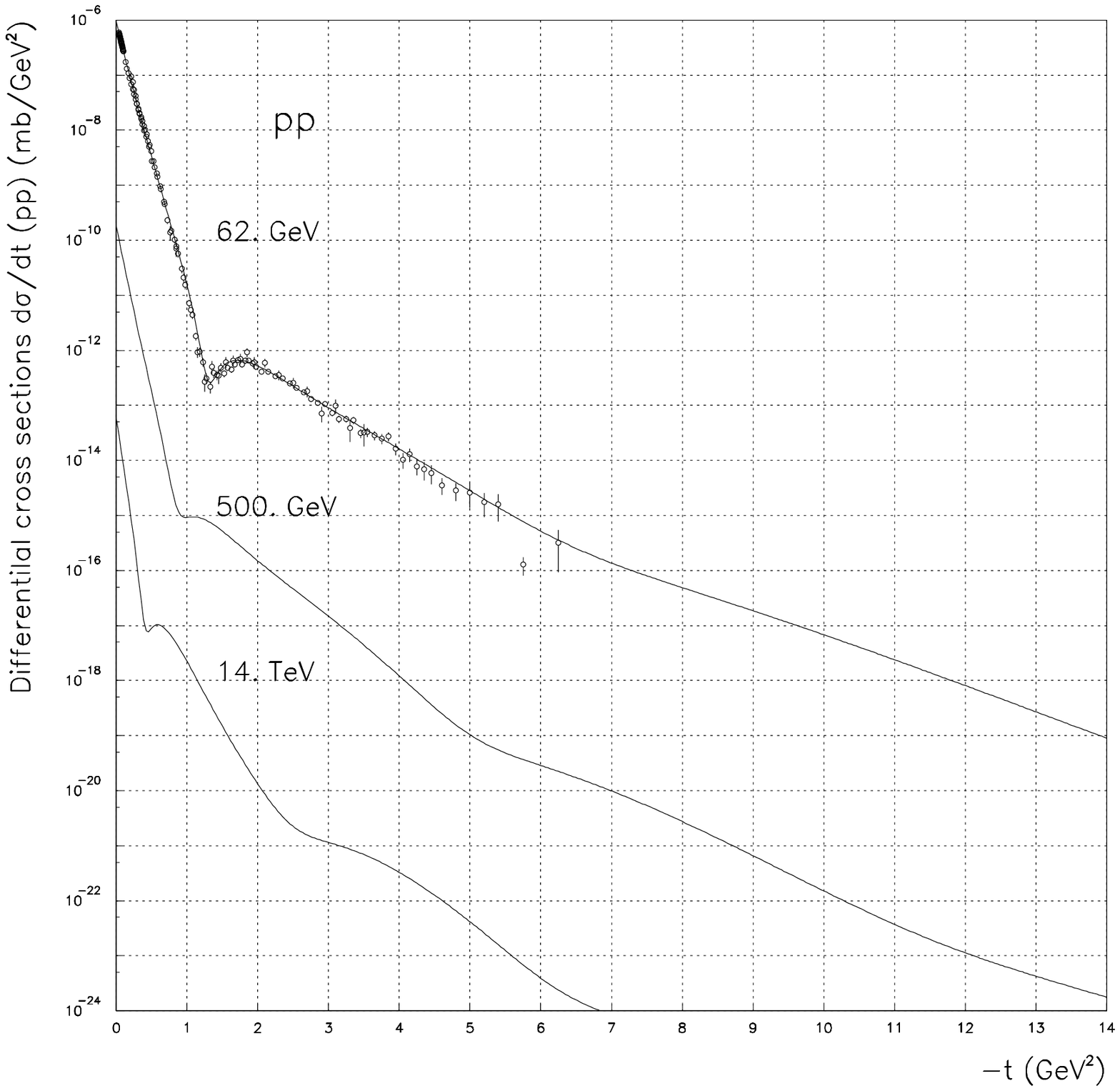}}
\vskip -3.cm
\caption{Predictions of the model for the differential cross-section of 
$p p$ scattering 
which will be mesured at LHC with $\sqrt{s}=14.\; TeV$ and at 
RHIC $\sqrt{s}=500.\; GeV$. The data 
corresponding to the energy $\sqrt{s}=62.\; GeV$ is multiplied by
 $10^{-8}$, RHIC by $10^{-12}$, and that of LHC by $10^{-16}$.
\label{fig:difpplhc}}
\end{figure}

The results are shown in fig.
\ref{fig:tot}, \ref{fig:elastic}, \ref{fig:reim}, \ref{fig:ppdif},
\ref{fig:ppdif1}, \ref{fig:difpbarp}, \ref{fig:difpbarp1},
\ref{fig:difpplhc}. 

We do not include elastic cross-section data sets into the fit and
predictions of the model for elastic
cross-sections can be seen in fig.~\ref{fig:elastic}. 

\begin{figure}[H]
\centering
\vspace*{-2.cm}
\epsfxsize=70mm \epsffile{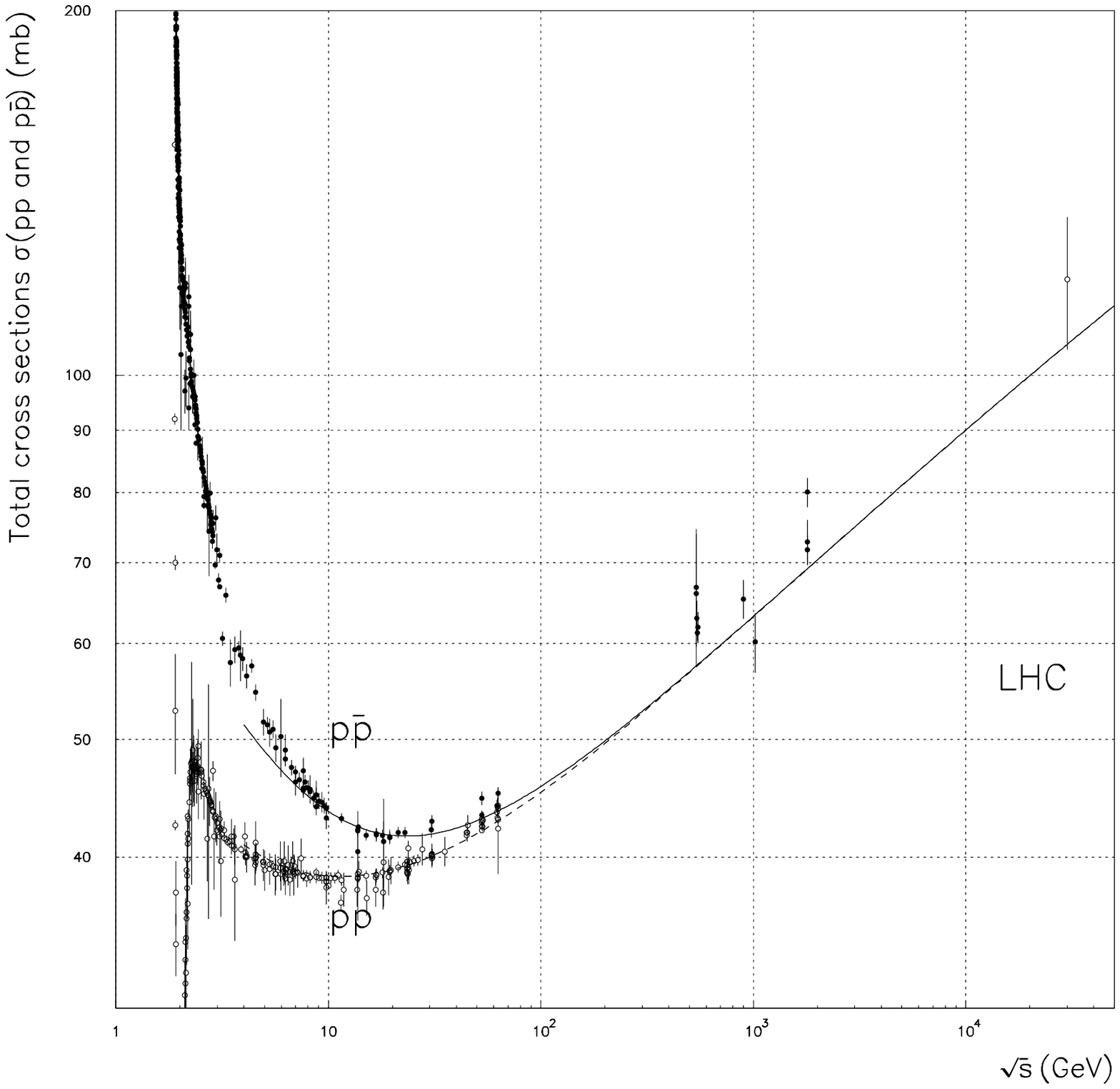}
\vspace*{-1.2cm}
\caption {Total cross sections of $pp$ scattering 
(hollow circles)  and $\bar p p$ scattering (full circles)
and curves corresponding to their description in the two-Pomeron model.
\label{fig:tot2p}}
\vspace*{-0.5cm}
\epsfxsize=70mm \epsffile{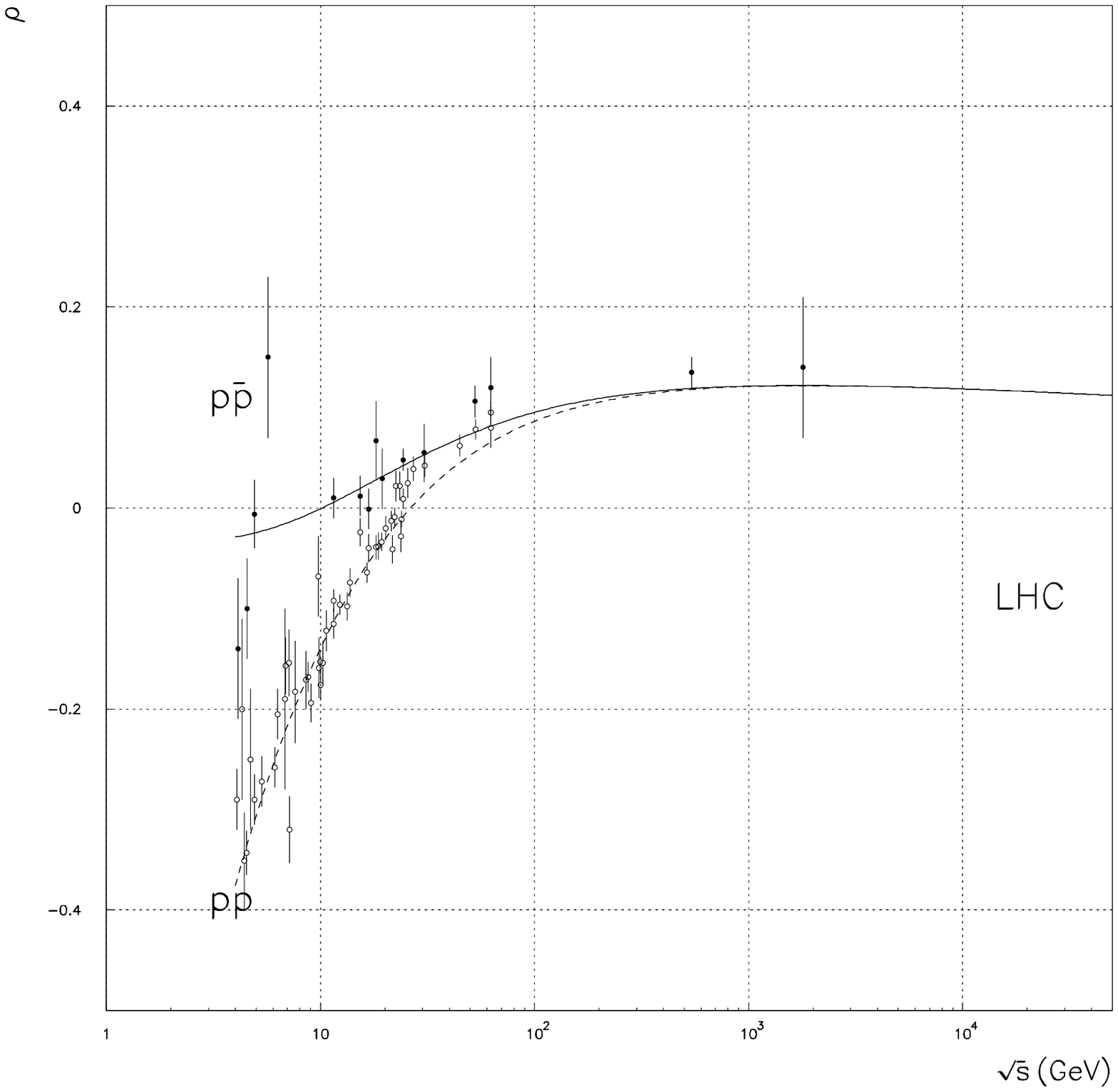}
\vspace*{-1.2cm}
\caption{
Ratios of the real to the imaginary part of the forward $pp$ scattering amplitude 
(hollow circles)  and $\bar p p$ scattering  amplitude (full circles)
and curves corresponding to their description in the two-Pomeron model. 
\label{fig:reim2p}}
\end{figure}

\begin{figure}[H]
\centering
{\vspace*{ -2cm} \epsfxsize=70mm \epsffile{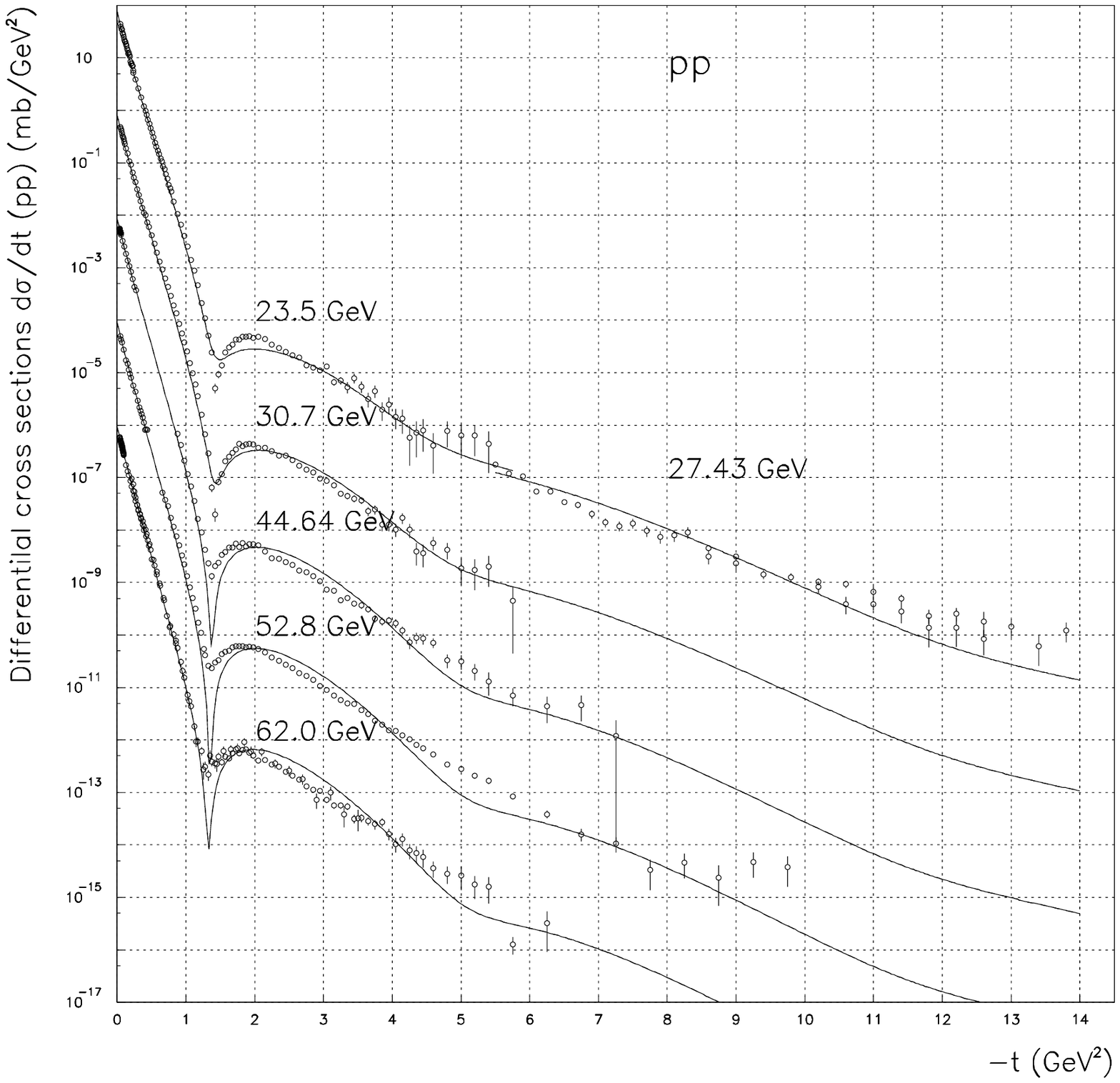}}
\vskip -1.2cm
\caption{Differential cross-sections for $pp$ scattering
and curves corresponding to their description in the two-Pomeron model. 
A $10^{-2}$ factor between each successive set of data is omitted. 
\label{fig:ppdif2p}}
{\vspace*{ -0.5cm} \epsfxsize=70mm \epsffile{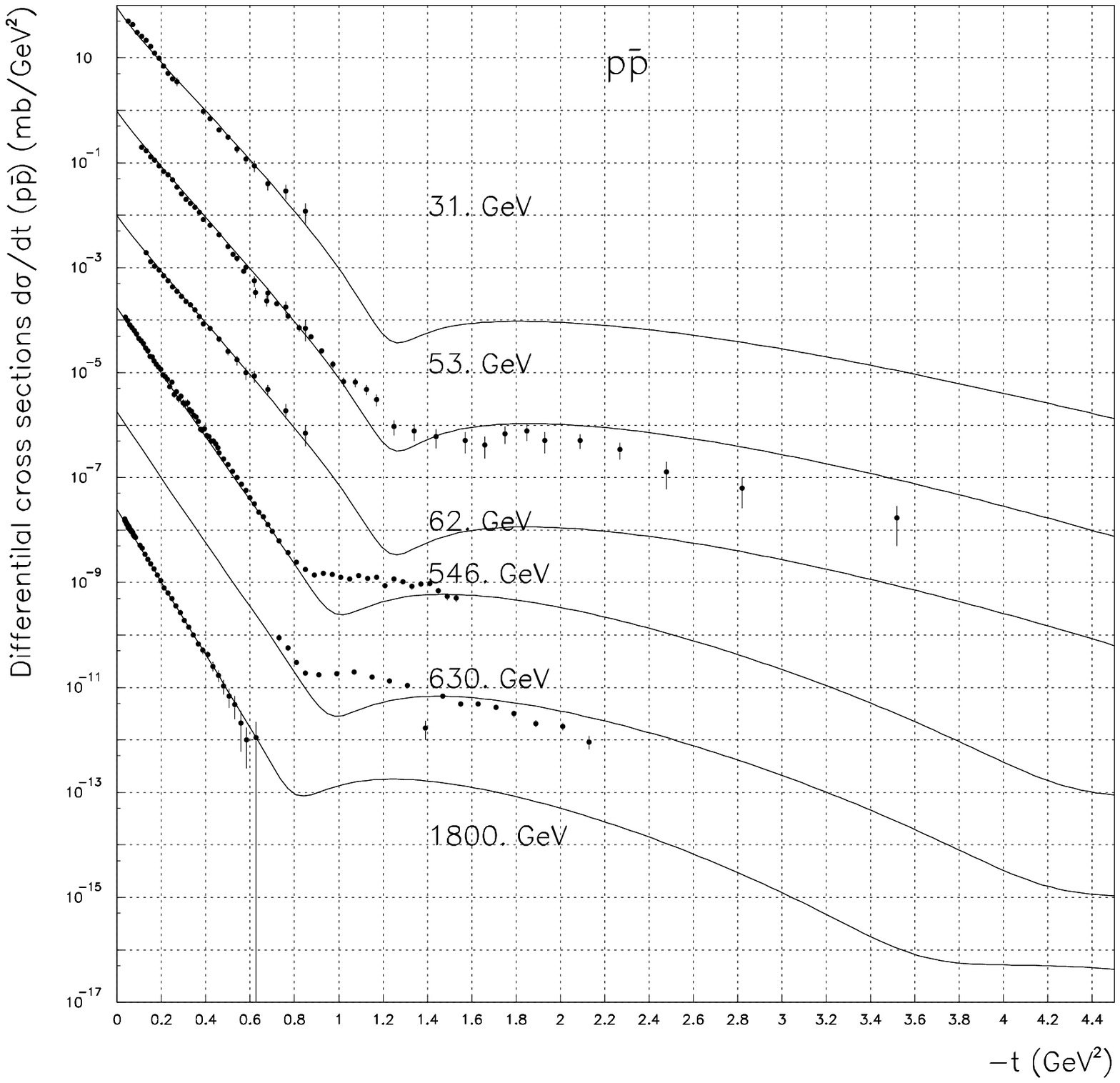}}
\vskip -1.2cm
\caption{Differential cross-sections for $\bar p p$ scattering
and curves corresponding to their description in the two-Pomeron model. 
A $10^{-2}$ factor between each successive set of data is omitted. 
\label{fig:difpbarp2p}}
\end{figure}

\begin{figure}[H]
{\vspace*{ -2cm} \epsfxsize=140mm \epsffile{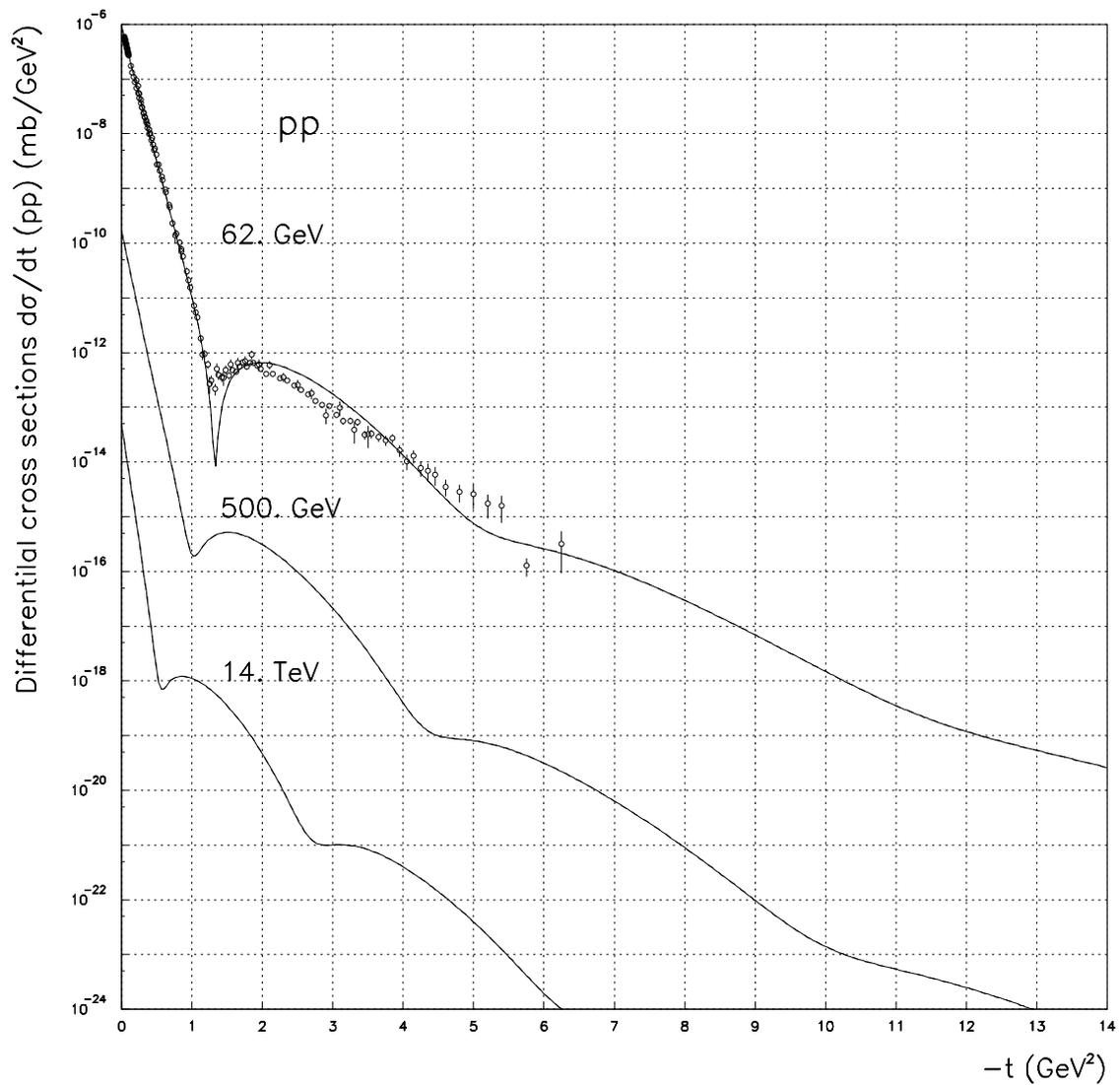}}
\vskip -3.cm
\caption{Predictions of the two-Pomeron model for the 
differential cross-section of 
$p p$ scattering 
which will be mesured at LHC with $\sqrt{s}=14.\; TeV$ and at 
RHIC $\sqrt{s}=500.\; GeV$. The data 
corresponding to the energy $\sqrt{s}=62.\; GeV$ is multiplied by
 $10^{-8}$, RHIC by $10^{-12}$, and that of LHC by $10^{-16}$.
\label{fig:difpplhc2p}}
\end{figure}

It is instructive to compare these results with a two-Pomeron option.
We give corresponding results in fig.~\ref{fig:tot2p}, 
\ref{fig:reim2p}, \ref{fig:ppdif2p},
 \ref{fig:difpbarp2p}, \ref{fig:difpplhc2p}. In this
case $\chi^2/{\rm d.o.f.}=10.87$ what can be considered as an argument in favour of our
hypothesis of many Pomerons.
The parameters obtained in the two-Pomeron 
option are presented in Table~\ref{tab:3} (all the errors are obtained 
according to MINUIT output).
\begin{table}[H]
\begin{center}
\begin{tabular}{|l|l|l|l|}
\hline
& {\bf Pomeron$_{\bf 1}$} & & {\bf $\bf f$-Reggeon}  \\
\hline
$\Delta_{{\Bbb P}_1}$ & $0.0859\pm 0.0021 $ & $\Delta_{f}$ & $-0.31$ (FIXED) \\
$c_{{\Bbb P}_1}$ & $53.18\pm 0.86$  & $c_{f}$ & $188.51\pm 12.13$            \\
$\alpha'_{{\Bbb P}_1}$& $0.360\pm 0.009\;(GeV^{-2})$  & $\alpha'_{f}$& $0.84\;(GeV^{-2})$ (FIXED)    \\
$r^2_{{\Bbb P}_1}$& $9.595\pm 0.6289\;(GeV^{-2})$&$r^2_{f}$ & $41.424\pm 7.971\;(GeV^{-2})$  \\
\hline
& {\bf Pomeron$_{\bf 2}$}& & {\bf $\bf \omega$-Reggeon}  \\
\hline
$\Delta_{{\Bbb P}_2}$ & $0.14437\pm 0.0051$  & $\Delta_{\omega}$ & $-0.53$ (FIXED)   \\
$c_{{\Bbb P}_2}$ & $6.87\pm 0.36$    & $c_{\omega}$ & $-171.36\pm8.23 $           \\
$\alpha'_{{\Bbb P}_2}$& $0.082\pm 0.004\;(GeV^{-2})$    & $\alpha'_{\omega}$& $0.93\;(GeV^{-2})$ (FIXED)     \\
$r^2_{{\Bbb P}_2}$& $4.765\pm 0.2533\;(GeV^{-2})$ &$r^2_{\omega}$ & $2.621\pm6.362\;(GeV^{-2})$  \\
\hline
& {\bf Odderon}& & \\
\hline
$\Delta_{{\Bbb O}}$ & $-0.2707\pm 0.1178$  &$s_0$& $1.0\;(GeV^2)$ (FIXED)   \\
$c_{{\Bbb O}}$ & $1.8134\pm1.4837$      & &        \\
$\alpha'_{{\Bbb O}}$& $0.029\pm 0.023\;(GeV^{-2})$   & &     \\
$r^2_{{\Bbb O}}$& $1.159\pm0.591\;(GeV^{-2})$ & & \\
\hline
\end{tabular}
\end{center}
\caption{Parameters obtained by fitting to the data with two
Pomeron contributions. \label{tab:3}}
\end{table}





\section{CONCLUSION AND DISSCUSSION}
Above we have developed a model	which is based on a general argument of multiplicity of
the Pomeron Regge poles in the eikonal.
The present model shows a very good descriprion of the available data
for all momenta ($0.01\le |t|\le 14.\; GeV^2$) and energies 
($8.\le\sqrt{s}\le 1800.\; GeV$) so that $\chi^2/{\rm d.o.f.}=2.74$. 

The model predicts the appearance of two dips in the differential cross-section 
which will be measured at LHC, fig.~\ref{fig:difpplhc}, and this
prediction is stable in the sence that all two models with
two and three Pomeron contributions predict the same behaviour
of the differential cross-section with two dips. These dips are
to apear in the region $t_1\simeq -0.5\; GeV^2$ and $t_2\simeq -2.5\; GeV^2$
which is in agreement with other predictions 
(model~\cite{multipomeronPredazzi}). 

We predict the following values of the total cross-section, elastic cross-section, and the ratio of 
 real to imaginary part of the amplitude for the LHC:
\bea
\nonumber
\sqrt{s}=14.\; TeV \;, \\
\nonumber
\\
\sigma_{tot}^{pp} = 106.73\;\;(mb)\;_{-\; 8.50mb}^{+7.56 \;mb}\;, \\
\nonumber
\\ 
\nonumber
\sigma_{elastic}^{pp} = 29.19\;\;(mb)\;_{-2.83\; mb}^{+3.58 \;mb}\;, \\
\nonumber
\\
\nonumber
\rho^{pp} = 0.1378\;_{-0.0612}^{+0.0042}\;.
\eea

Predictions for RHIC are:
\bea
\nonumber
\sqrt{s}=500.\; GeV \;, \\
\nonumber
\\
\sigma_{tot}^{pp} = 59.05\;\;(mb)\;_{-3.10\; mb}^{+2.94 \;mb}\;, \\
\nonumber
\\
\nonumber
\sigma_{elastic}^{pp} = 12.29\;\;(mb)\;_{-0.76\; mb}^{+0.79 \;mb}\;, \\
\nonumber
\\
\nonumber
\rho^{pp} = 0.1327\;_{-0.0071}^{+0.0052}\;.
\eea

  The parameters of the Pomeron trajectories are:
\bea
\nonumber
\alpha(0)_{{\Bbb P}_1}=1.058,\;\;\alpha'(0)_{{\Bbb P}_1}=0.560\;(GeV^{-2}); \\
\alpha(0)_{{\Bbb P}_2}=1.167,\;\;\alpha'(0)_{{\Bbb P}_2}=0.273\;(GeV^{-2}); \\
\nonumber
\alpha(0)_{{\Bbb P}_3}=1.203,\;\;\alpha'(0)_{{\Bbb P}_3}=0.094\;(GeV^{-2}). 
\eea

Their coupling constants fulfil the following inequality:
\be
c_{{\Bbb P}_1}> c_{{\Bbb P}_2}> c_{{\Bbb P}_3}
\ee
 
It may mean that there exists a series of Pomeron contributions each term 
having the form~(\ref{eq:eikonalform}), or these three contributions effectively
emulate one nonlinear Pomeron trajectory.

\begin{figure}[H]
\centering
\parbox[t]{6.cm}{\vspace*{ 0cm} \epsfxsize=60mm \epsffile{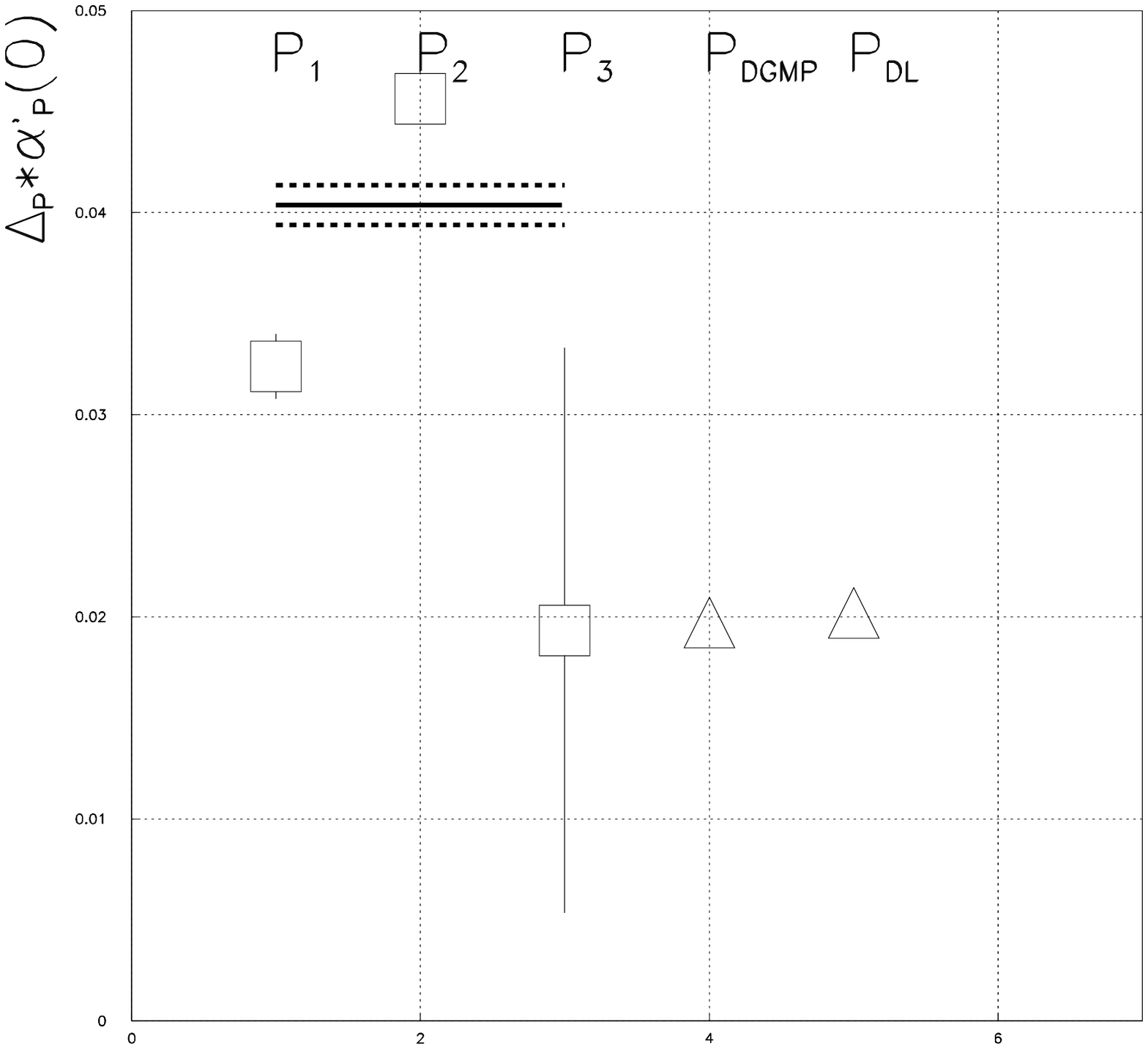}}
\hfill~\parbox[t]{6.cm}{\vspace*{ 0cm} \epsfxsize=60mm \epsffile{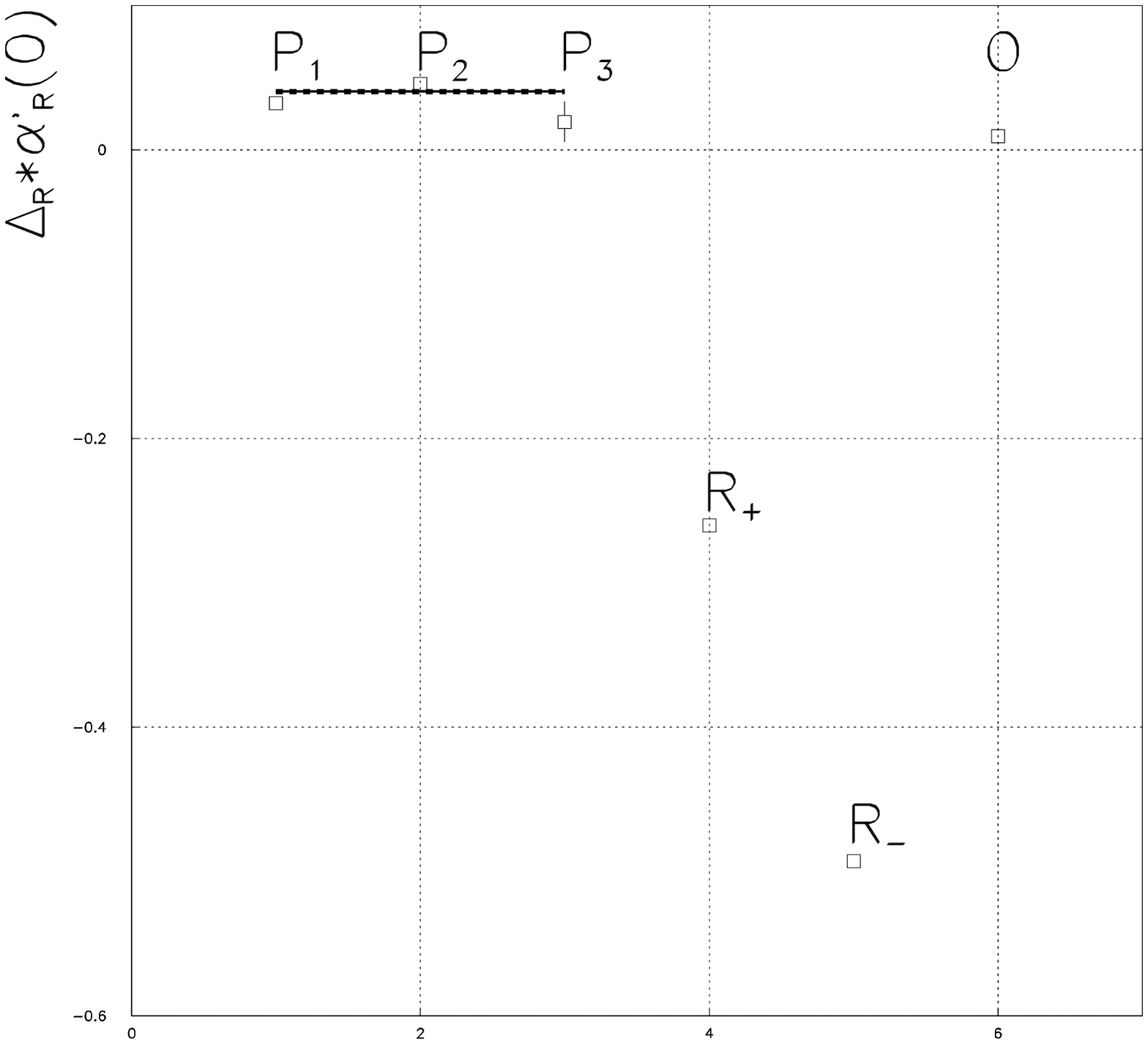}}
\vskip -1.cm
\parbox[t]{6.5cm}{\caption{Products of intercepts to slopes for the Pomerons (hollow squares) and the Pomeron in a generalized eikonalization model ${\Bbb P}_{DGMP}$~\cite{multipomeronPredazzi}, Donnachie\&Landshoff Supercritical Pomeron ${\Bbb P}_{DL}$~\cite{donnachie} (hollow triangles). Solid line corresponds to the mean value of the product (for the three Pomerons) and dashed lines correspond to its error corridor.
\label{fig:interslope}}}
\hfill~\parbox[t]{6.cm}{\caption{Products of intercepts to slopes for the Pomerons, the Odderon, and Reggeons used in the present model.\label{fig:interslope1}}}
\end{figure}

The intercepts and slopes fulfil the following inequalities:
\bea
\Delta_{{\Bbb P}_1}< \Delta_{{\Bbb P}_2}< \Delta_{{\Bbb P}_3} \\ \nonumber
\alpha'_{{\Bbb P}_1}(0)> \alpha'_{{\Bbb P}_2}(0)> \alpha'_{{\Bbb P}_3}(0),
\eea
i.e. the higher is the intercept the lower is the slope.
We observe that the product of the intercept and the slope is approximately the same 
for all the  Pomerons, 
$\Delta\cdot\alpha'(0)\simeq 0.040\pm0.0009\; (GeV^{-2})$. This is seen in fig.~\ref{fig:interslope}.
This constant seems suprisingly universal if compared with the products of other Reggeon 
parameters used in this model (fig.~\ref{fig:interslope1}). At present we have no clear 
understanding of this universality.

We can only remind that high energy asymptotic behaviour of total and elastic 
cross-sections in Regge-eikonal approach have the following form:
\bea
\sigma_{tot}(s)\Big|_{s\rightarrow\infty} \rightarrow 8 \pi \alpha_{\Bbb P}'(0) \Delta_{\Bbb P} \ln^2 (s/s_0)
\;,\\ \nonumber
\sigma_{elastic}(s)\Big|_{s\rightarrow\infty} \rightarrow 4 \pi \alpha_{\Bbb P}'(0) \Delta_{\Bbb P} \ln^2 (s/s_0)\;,
\eea 
and the constant $\alpha_{\Bbb P}'(0) \Delta_{\Bbb P}\; (GeV^{-2})$ (there $\Bbb P$ stands for 
the rightmost singularity of the eikonal function in $J$--plane) 
defines a universal (independent on colliding beams) asymptotic
behaviour.

It is interesting to enlist the following characteristic properties 
of the Pomerons used in this paper.

The first of the Pomerons (`$Pomeron_1$') possesses the properties that we expect
from the string picture~\cite{Soloviev} of Reggeons, 
i.e. $\alpha'(0)_{{\Bbb P}}=\frac{1}{2}\alpha'(0)_{f}=0.42\;(GeV^{-2})$ 
and indeed $\alpha'(0)_{{\Bbb P}_1}=0.559\pm 0.078\;(GeV^{-2})$.

The second Pomeron (`$Pomeron_2$') is close to what is called ``supercritical Pomeron"
with the slope $\alpha'(0)_{{\Bbb P}_2}=0.273\pm0.005\;(GeV^{-2})$ close to its ``world'' value $\alpha'(0)_{{\Bbb P}}\simeq 0.25\;(GeV^{-2})$.
 
The third Pomeron (`$Pomeron_3$') is reminiscent of what is known as a ``hard'' 
( or perturbative QCD )Pomeron.
Its parameters 
($\alpha(0)_{{\Bbb P}_3}=1.203$, $\alpha'(0)_{{\Bbb P}_3}=0.094\;(GeV^{-2})$)
are close to the calculated parameters of the perturbative Pomeron, 
which arise from the summation of reggeized gluon ladders 
and BFKL equation~\cite{bfkl}: $\alpha(0)_{{\Bbb P}}^{BFKL}\simeq 1.2,\;\;\alpha'(0)_{{\Bbb P}}^{BFKL}\sim 0.\;(GeV^{-2})$. 
The fact of arising of a ``hard" Pomeron in a presumably ``soft" framework can seem quite unexpected.
However we are not particularly inclined to identify straightforwardly 
``our hard Pomeron" with that
which is a subject of perturbative QCD studies.

The Odderon has the following parameters: $\alpha(0)_{\Bbb O}=1.192,\;\;\alpha'(0)_{{\Bbb O}}=0.048\;(GeV^{-2})$ in agreement with unitarity constraints 
Eq.~(\ref{eq:uconstraints}).
The Odderon intercept is positive and close to that of the $Pomeron_3$. The slope
is almost zero. The coupling is so small that only high-t data may be sensible
to the Odderon contribution.

Assuming that one can neglect the non-linearities of Regge trajectories and making use of a simple
parametrization
\be
\alpha(m^2)=\alpha(0)+\alpha'(0)\cdot m^2\;,
\ee
we can try to estimate the corresponding spectroscopic content of our model.

Then $\Re {\rm e}\alpha(m^2)=J$, where J is an integer number corresponding 
to the spin of a particle which we should find lying on the trajectory. 

\begin{figure}[H]
\centering
{\vspace*{ -2cm} \epsfxsize=100mm \epsffile{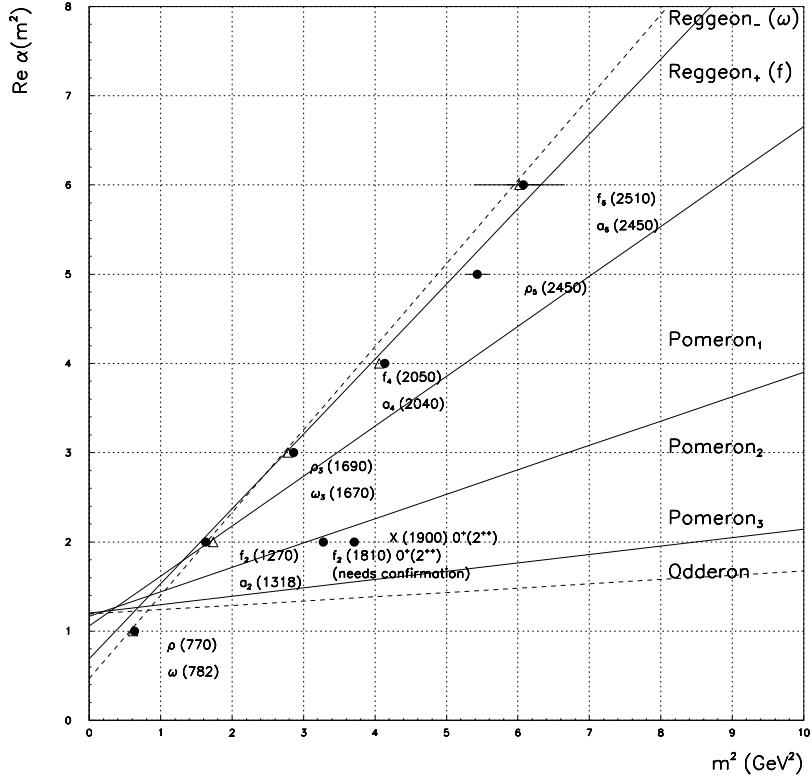}}
\vskip -2.cm
\parbox[t]{10.cm}{\caption{Regge trajectories of secondary Reggeons, three Pomerons and the Odderon.
\label{fig:traject}}}
\end{figure}

The trajectories are depicted in fig.~\ref{fig:traject}. The $C+$ Reggeon trajectory is in fact a combination of two families of mesons $f$ and $a_2$. 
The $C-$ Reggeon trajectory is a combination of two families of mesons $\omega$ and $\rho$. As is seen, the secondary Reggeon trajectories fairly well describe the spectrum of mesons.

Among the mesons with appropriate quantum numbers there exesit two that fit
the Pomeron trajectory ($0^+J^{++}$): $f_2(1810)\;\;0^+2^{++}$ with mass $m=1815\pm12 \;\; MeV$ and $X(1900)\;\;0^+2^{++}$ with mass $m=1926\pm12 \;\; MeV$
. One of them is supposed to be on $Pomeron_2$ trajectory.

\section*{ACKNOWLEDGMENTS}
We would like to thank professor Enrico Predazzi for useful discussions
and for reading the manuscript.


\end{document}